\documentclass[showpacs,preprintnumbers,amsmath,amssymb]{revtex4}
\usepackage{graphicx}
\usepackage{dcolumn,amssymb, latexsym, amsmath, eucal}

\def\gr{general relativity }

\def\qm{quantum mechanics }

\def\bs{\hspace{-5pt}}

\begin{document}

\title{{\huge Why Eppley and Hannah's Experiment Isn't}}

\author{James Mattingly}
 \email{jmm67@georgetown.edu}
\affiliation{Georgetown University}%

\date{\today}

\begin{abstract}
It is shown that Eppley and Hannah's thought experiment establishing that gravity must be quantized is fatally flawed.  The device they propose, even if built, cannot establish their claims, nor is it plausible that it can be built with any materials compatible with the values of $c, \hbar, G$.  Finally the device, and any reasonable modification of it, would be so massive as to be within its own Schwarzschild radius---a fatal flaw for any thought experiment. 
\end{abstract}

\maketitle
\section*{Introduction}

Physicists (and philosophers and other lay readers of physics) have by and large accepted claims to the effect that the gravitational field must be quantum mechanical in nature.  Remarkably, the evidence for these claims is extremely thin, and indeed what there is is extremely dubious.  For general analyses of the failure to establish the case for quantization, see Mattingly's discussion\cite{jm1, jm2}, and Calender and Huggett's editorial introduction\cite{ch} to their (2001).

Even though there is little experimental verification that the gravitational field is quantized, it seems that both physicists and philosophers take very seriously two separate experiments that apparently establish that gravity is quantum mechanical.   Widely cited in discussions of the necessity for quantizing gravity are Eppley and Hannah's\cite{eh} and Page and Geilker's\cite{pg} experiments.  While Page and Geilker do carry out their experiment, the significance of their result was called into question even before the experiment was performed, by the very person (Kibble \cite{k}) who suggested the experiment in the first place.  I will not here discuss their experiment since its significance has been decisively undermined\cite{jm1, jm2, ch, k}, and because it applies only to one specific version of semiclassical gravity, semiclassical general relativity.  

While Eppley and Hannah's experiment hasn't been performed the issue of its significance is more delicate.  They proposed a brilliant thought experiment to demonstrate the invalidity of {\em any} non-quantum mechanical version of gravitation theory that is general relativistic in non-quantum limit.  Their detailed analysis is meant to establish that {\em any} non-quantized gravitational theory is inconsistent with either the first signal principle of special relativity, momentum conservation, or Heisenberg's uncertainty principle.  This would be a profound result.  I will show however that their experiment cannot be carried out.  In particular, in any experimental situation suitable to producing the results they require, the device they use to measure these results cannot be built {\em in principle}.  We might offer therefore, in analogy with cosmic protection hypotheses that there are no naked singularities, the semiclassical protection hypothesis that possible inconsistencies in semiclassical gravity are hidden from observation.

The result here presented is important because Eppley and Hannah's paper in particular has had a significant impact on all future discussions of the question of quantizing the gravitational field.  There are few discussions of the evidence that gravity is quantized that do not appeal to Eppley and Hannah's result, and many take their experiment to have settled the question.  Here I will not directly challenge the view that the gravitational field really is quantum mechanical.  Instead I will merely show that one important pillar supporting that view is without foundation.  One important reason to reconsider this experiment is that it gives a misleading picture of what stands in the way of a non-quantized theory of gravity.  Showing that Eppley and Hannah's method fails may prompt the development of a different, more successful experiment.  And that experiment might itself be a useful pointer toward quantum gravity.

\section*{Eppley and Hannah's Thought Experiment}

\vspace{2em}

In 1975  Eppley and Hannah\cite{eh} proposed their thought experiment to show 
that the gravitational field must be quantized.  The experiment uses a gravity wave to measure 
the position and momentum of a macroscopic body such that $\Delta 
p_{x}\Delta x < \hbar$ thus violating the Heisenberg uncertainty 
principle.  The key idea is that a classical wave may have arbitrarily 
low momentum and, simultaneously, arbitrarily short wavelength.  To find a 
conflict with quantum mechanics they couple a short 
wavelength/low momentum gravitational wave to a quantum system.  The 
gravitational wave may then be used to localize a particle within one wavelength 
while introducing vanishingly small uncertainty into the particle's 
momentum.

Eppley and Hannah are well aware that a thought experiment must be in some sense
realistic, and they adopt explicit standards:  While the experiment may never be carried out---and
indeed we may never even possess the engineering skill or the raw
power to carry it out---it still must be physically possible to do so
within the bounds of the theory employed for the description of the
device---in this case the semiclassical theory of gravity.  They say, 

\begin{quotation}

\noindent We want to show that the experiment is possible in principle, in the sense that it does not require any masses, lengths, or times greater than those of the universe.(64)

\end{quotation}

\noindent
and they conclude

\begin{quotation}
    
    \noindent
    Our experiment is fantastically difficult to perform, but
    nevertheless in principle possible.(67)

\end{quotation}

I agree with Eppley and Hannah that they need merely show that their experiment is possible in principle.  I will adopt their standard---that they need to show that no ``masses, lengths, or times greater than those of the universe" be required.(64)  Still I claim that their experiment {\em cannot} be performed even in principle.

We must also keep in mind that the burden of proof lies squarely on Eppley and Hannah.  Proposing a thought experiment requires experimental controls at least as stringent as those on an ordinary experiment.  It will not do to say what the experiment would have shown were it possible to construct in principle.  It must be shown that the experiment really is possible in principle.  For well established physical principles, such claims can be established quite casually.  If, for example, a thought experiment requires that the some structure be imagined to have been built according to standard construction techniques, then we needn't demand an exhaustive demonstration that the structure is possible in principle.  But for more exotic claims we do.  If, for example, an experiment requires some exotic material, we must demand that it be demonstrated that such materials can be found given our current understanding of the laws of physics.  Suppose we need a stable lump of some quantity of radioactive material.  We must then show that that amount is less than the critical mass for that material.

It is not necessary to show that no thought experiment.  Instead we need only show that Eppley and Hannah have not delivered on their promise to produce such an experiment, or even one that requires only trivial and easily generated modifications.  What follows will establish that Eppley and Hannah have not made good on their claim that their experiment can be performed.  Thus one of the best arguments for the quantization of gravity fails.

I proceed in two stages.  In the first stage I show that even if there {\em were} a device that succeeded in measuring what they propose, that would still not show that gravity must be quantized.  In the second stage I show that their device {\em cannot} be made to work.

\subsection{Their Argument}

I'll consider the construction of the device shortly.  First I lay out
their argument.  Beginning with a poorly localized particle of sharply defined momentum
Eppley and Hannah consider a gravitational wave scattering from the
particle.  They claim that there are two exhaustive and mutually
exclusive cases.

\begin{quotation}
    
    \noindent Either the scattering event constitutes a position measurement of
    the particle with a consequent collapse of its position wave
    function to a smaller region of localization, or else it is not a
    position measurement and does not result in collapse. 
    (1977, 53)
    
\end{quotation}

They argue that each of these cases will conflict with 
well-entrenched physical principles, and hence that the coupling of a 
classical field to a quantum field is impossible.  They take it for 
granted that the ``first signal'' principle of special relativity 
holds.  That is, they claim that signalling ``across arbitrary 
spacelike distances \ldots clearly violates relativity.''(57)  
Likewise they appeal to conservation of momentum and to the 
Heisenberg uncertainty principle.  There are two cases to consider.

\subsubsection{Case 1}

Suppose that the gravitational wave does collapse the particle's
wavefunction.  Then, Eppley and Hannah argue, the particle is now
localized within one wavelength of the gravitational wave.  By
determining the point of interaction, we can attribute a sharply
defined position to the particle and, since we introduced arbitrarily
little momentum disturbance into the system it should still have a well-defined momentum.  We are presented
with two sub-cases: either the interaction conserves momentum but the
uncertainty principle is violated (since now the particle has well-defined momentum and position), or if the uncertainty principle is not violated then the interaction does not conserve momentum.

Of the latter sub-case they say, ``If the classical probe gives the
particle a very good position localization, then quantum mechanics
implies that the particle is now in a state of very high momentum.  If
the quantum description of the particle is valid, then momentum is not
conserved, since the momentum of the initial quantum state was very
well defined and the classical probe imparted negligible
momentum.''(54-55)  This argument has been challenged in a number of places\cite{jm1, jm2, ch}, and I will pursue it only briefly here, first noting that the statement of the uncertainty principle needs a slight modification.  What we can say is that the quantum description of matter implies that the uncertainty in the particle's momentum is now greater than it was, and thus the probability of finding it in a higher momentum state is greater.  So also is the probability of finding it in a lower momentum state.  This certainly does not imply that momentum is not conserved.

\vspace*{0.1in} 
\paragraph*{Subcase 1a.}\hfill\\

Quantum mechanics implies at most that the product of the
spread in values of the two observables has a definite lower
limit.  In other words, in this case we no longer can say, with great
precision, what the particle's momentum is.  It requires another
argument to tell us if momentum is conserved in the interaction.  We did not really prepare our particle in an eigenstate of
momentum, with $\Delta p = 0$.  Rather we prepared the particle with a
sharply peaked, but finitely spread, momentum distribution centered
about the measured value.  How is this significant?  Even in its
initial momentum state, there is a finite probability of
scattering into any other momentum state \emph{in the absence of
any outside interaction}.  How does this work?  The probability of
observing momentum within a small interval $d$ (the resolution of our
detector) of $p' \neq p$ is given by:
\begin{equation*}
    P_{\phi(p')} \approx |\phi(p')|^{2}d.
\end{equation*}
Unless the wavefunction has zero support outside some small range of
values, this probability will be non-zero.  So to decide whether momentum
is conserved, we need to do more than show that, for this experiment,
the initial measured momentum of the system is possibly different from
its final measured momentum.  

Now we would have a good case that momentum is not conserved if the
\emph{expectation value} of the momentum changed during the
measurement.  I will assume, for simplicity, that the initial distribution
is a Gaussian of width $d$ peaked about $p$, the original observed
momentum.  The conclusion of my present 
argument---that a change in the distribution of momentum need not 
indicate a change in the expectation value of momentum---holds for 
general distributions.  Eppley and Hannah claim that if the
uncertainty principle is obeyed then the momentum must change.  I
have indicated the most charitable and plausible way I can see to read
this claim---the expectation value of the momentum must change if the
the uncertainty principle is not violated.  The wavefunction for a
momentum space Gaussian is:
\begin{equation*}
    \phi(p) = \sqrt{\frac{d}{\hbar\sqrt{\pi}}}exp(\frac{-(p -\hbar k)^{2}d^{2}}{2\hbar^{2}}).
\end{equation*}
Here k is the wavenumber and d is the width of the Gaussian.  The
respective expectation values of the momentum for the new and old wave
functions are:
\begin{equation*}
\langle p(\phi)\rangle =\hbar k \hspace{1.0em}\mathrm{and}\hspace{1.0em} \langle p(\phi')\rangle =\hbar k'.
\end{equation*}

What can we say about the relation between $k$ and $k'$?  This isn't at all clear 
from what has been said so far, but in effect their relation is the 
relation between the new and old momentum.  But what is that 
relation?  Eppley and Hannah's argument relies on the claim that, in 
order to obey the uncertainty principle, $k$ must differ from $k'$ by 
more than the momentum (divided by $\hbar$) imparted by the classical 
gravitational wave.  This claim is false.  All that the uncertainty 
principle requires is that the width of the Gaussian that modifies 
the momentum space wave function must change.  This is the factor $d$ 
in the Gaussian wave packet above.  It is thus perfectly consistent 
with quantum mechanics to assume that, neglecting the small momentum 
transfer about which we aren't concerned,
\begin{equation*}
\langle p(\phi)\rangle =\langle p(\phi')\rangle.
\end{equation*}

Thus we have to show something more complicated than the fact that
after the gravitational interaction the particle may now be found in a
very different momentum state.  Perhaps we could instead show that
energy is not conserved during the measurement.  Calendar and Huggett 
\cite{ch} consider this possibility at some length.  Roughly the idea is
that energy conservation during measurement is a problem for
\emph{any} collapse interpretation.  Specifically the brief suspension
of the Schr\"{o}dinger equation during measurement implies that we
should have very little confidence, during measurement, in any
conservation law derived from this equation.  (As opposed to our
experimentally very well justified confidence in energy being
conserved statistically.) But whatever one's feeling about the
viability of energy conservation for collapse interpretations, it
should be clear that nothing in Eppley and Hannah's argument implies
that energy conservation fails for their experiment, since kinetic
energy follows momentum.  To show that energy conservation is violated would require showing that its expectation value changes, and we are offered no independent grounds for that.  The lesson here is that merely causing the
width of a distribution to change has no bearing on the location of
the mean of the distribution.  At the very least, a great deal more
argument is required before we can accept Eppley and Hannah's claim
that ``[i]f the classical probe gives the particle a very good
position localization, then quantum mechanics implies that the
particle is now in a state of very high momentum.''(154) Nothing in
their analysis supports this claim, and my analysis (under what seems
to be the only plausible reading of that claim) of what quantum
mechanics requires strongly supports its denial.

I will therefore assume that if the gravitational wave does collapse the particle's wave function it localizes it will preserving its sharply defined momentum state.  Would this really 
violate the uncertainty principle?

\vspace*{0.1in} 
\paragraph*{Subcase 1b.}\hfill\\

The uncertainty principle is sometimes misleadingly presented as a
feature of individual particles.  The claim is that any measurement of
the position of a particle must introduce uncontrollable changes in its
momentum.  But, there are empirically adequate interpretations of
quantum mechanics for which these relations are epistemological and
not a fundamental feature of the world.  On the de Broglie-Bohm theory
position and mechanical momentum are always well defined. 
It is not inconceivable on such an interpretation that one could find a way to measure them
together with arbitrary accuracy, but only that a purely quantum mechanical measurement could not.  Given that the viability of a hybrid theory is at issue, we cannot assume that such measurements are impossible.  Yet the de Broglie-Bohm interpretation is empirically equivalent to the standard, Copenhagen interpretation.  Similarly Everett's relative state formulation of quantum mechanics is empirically identical to the Copenhagen interpretation, and particles there always have well-defined position and momentum.  Thus we cannot assume that sharply localizing a particle simultaneously in both momentum and position space is inconsistent with the uncertainty principle.  

It is therefore more proper to take
the uncertainty principle as a statement of distributions of
measurement outcomes.  Only the preparation of an ensemble of such particles that, as a
group, violate the uncertainty principle should cause concern. 
But Eppley and Hannah's thought experiment---even if it were to
work---could not be used do this since they only have the resources to prepare one particle at a time.  To violate the uncertainty
principle would require an interference experiment of some kind---like
the two-slit experiment.  Eppley and Hannah implicitly agree with this
view, claiming that the uncertainty relations are violated because ``a
beam of such particles sent through an arbitrarily narrow slit would
show no diffraction effects, contrary to fact.''(55)  But their
``particles'' have such small wavelengths with respect to their size,
that no interference experiment of that kind is possible.  Their ``arbitrarily narrow slit''
would have to be much narrower than the dimensions of the bodies, and therefore there could be no transmission through the device, much less scattering effects.  Perhaps it will be suggested that this
experiment could be done with smaller bodies that can, in principle, be diffracted. 
As far as I know, the largest bodies that can be diffracted have masses on the order of $100,000$ neutrons (a very complicated mesoscopic molecule).  Whether this is a matter of principle is unclear, but at least for the devices considered by Eppley and Hannah,  for
bodies of this mass, the radius of the Eppley and Hannah detection
device needs to be roughly $10^{59}cm.$ On the other hand, the radius
of the universe is of order $10^{28}cm.$  So even in principle we could not detect just
one, let alone an entire ensemble of such particles.   One might consider other ways to measure interference effects on macroscopic bodies.  Is that possible?  Without giving a detailed method that is possible in principle---the whole point of their article---we cannot conclude that it is.  I believe most would assume that, given the small value of $\hbar$, such experiments are impossible in principle.  In any case Eppley and Hannah have not shown that the uncertainty principle would be violated.

Even being able to interfere such macroscopic bodies is not enough.  For the experiment to show a violation of
the uncertainty principle, it is necessary that the position space
wave function not spread between the time that the particle is
localized and the time it is sent through the diffraction grating. That is to say that, if indeed we have produced particles with $\Delta p \Delta x \leq \hbar$, there is every reason to suppose that the wave-function will spread again, and this spreading occurs on very short time scales.  In fact, if we
have localized the particle so well that its position wave-function is
zero outside of some finite region, then the spreading occurs on
infinitesimal time scales.  Even for localization with only high
confidence---very low probability to find the particle outside the
region---the spread occurs rapidly enough that to ensure experimental
violation of the uncertainty principle would require an
additional measurement protocol unaddressed by Eppley and Hannah, in
order to assure that the spreading had no effect on their experiment.  The possibility of such an auxiliary protocol simply has not been established.  Normally of course one does not need to worry about such issues, but if we are assuming that the experiment only functions when the gravity wave collapses the wave function of the particle, then we must consider whether properties of that collapsed state can be maintained long enough to measure them.  In the context of a thought experiment, the burden is on the experimenter to show that all parts of the experiment, including this auxiliary component, are possible in principle.  Eppley and Hannah have not done so.  Absent a characterization of the experiment, and good reason to think it would have an outcome favorable to their claims, we haven't seen that well-established physical principles show that their experiment functions as advertised.

There are thus reasons, even in advance of a consideration of their device, to think that their experiment is in principle impossible.

\subsubsection{Case 2}

Suppose now that, rather than collapsing the particle's wavefunction,
the gravitational wave merely registers the presence of the particle's
wavefunction.  But a measurement of a particle's wavefunction in this
way would allow super-luminal signalling.  For before measurement, one
can split the position-space wavefunction of the particle into two
spacelike separated parts.  Then one
can measure the shape of the wavefunction in one region.  Depending
upon whether one finds no wavefunction, the full wavefunction, or half of
the wavefunction, one can tell whether someone else has either measured the
other part of the wavefunction (by normal quantum mechanical means)
and found the particle, measured and not found the particle, or not
made a measurement at all respectively.  Thus there is a prescription 
for super-luminal causation.  One can (since we are assuming here that 
collapse is instantaneous) tell instantly the state of the 
wavefunction in a spacelike separated region.

Notice again that Eppley and Hannah are working within the
``Copenhagen'' interpretation of quantum mechanics.  This implies
that they believe that measurement is accomplished by reduction of the
wave packet of a quantized particle.  So strictly speaking, their
argument has nothing at all to say to someone who subscribes to a
different interpretation of quantum mechanics.  

But, again on the de
Broglie/Bohm interpretation, there is never a ``collapse'' of the wave
function even for normal quantum measurements.  Similarly, it has become increasingly common in cosmological discussions to adopt some version of Everett's relative state description of quantum mechanics.  There again, all particles always have well-defined positions and momenta.  

\subsubsection{Conclusions from the argument.}

Going along with Eppley and Hannah and assuming that the  Copenhagen interpretation of quantum 
mechanics is correct, the most we can conclude from their argument is the claim 
that either non-quantized semiclassical gravity violates the first 
signal principle or a gravitational measurement does collapse the 
particle's wavefunction.  So even within the context of a Copenhagen interpretation of
quantum mechanics, and even if their device could be constructed,
their case that gravity must be quantized would still not be made.  
We would still need to make the case that the measurement will not 
collapse the wavefunction.  I assume in what follows then that we are operating under the conditions of {\em Case} $\mathit{1}$.  I leave here the discussion of Eppley and 
Hannah's argument and turn to a demonstration that, whatever the 
argument's merits, their experiment itself will not work even by {\em their own} 
standards.

\subsection{The Device}

I will level
four objections against their device.  One involving materials, one involving the temperature, one involving preparations, and the last involving gravity itself.  The first three are quite
serious and prevent their detector from functioning the way they claim
it should.  I doubt strongly that these objections could be overcome
in any modification of their device.  The fourth objection is damning,
and I show that any modification of their device, to mitigate the
effect of this problem, itself reintroduces the problem.  These
objections, and the conceptual problems I identified above, show that
their thought experiment cannot be considered persuasive evidence that
gravity must be quantized.

One might argue that the first three objections could be met in another universe.  That however would itself require significant argument.  Are the initial conditions of the universe contingent or lawlike?  I don't know, and nor does anyone else.  Ernst Mach's observation that the universe is only given once may not underwrite his sweeping rejection of any counterfactual claims about its initial state, but his observation is germane here nonetheless.  The issue is whether non-quantized gravity is possible in {\em this} universe, so any appeal to alleged facts about universes in general needs careful analysis that is beyond the scope of this note, and which is not undertaken in Eppley and Hannah's proposal.  No reason has been given to suppose that the existence of universes (spacetime manifolds) where semiclassical gravity is impossible has any bearing on the issue of its possibility simpliciter.  I am skeptical of such a conclusion especially since there are results that apparently establish the consistency of semiclassical gravity in $2$- and $3$-d spacetime---the only cases that are mathematically tractable.  (See for example Wald's (1994) discussion\cite{waldqft} and references therein.)

In any case, the fourth objection cannot be met in any universe with the same ratio of values of $c, G,$ and $\hbar$ as ours, and their relative values seem necessary for the existence of any stable matter at all.

Eppley and Hannah wish to localize sharply a mass that has a precisely 
defined momentum.  They do this by scattering a classical 
gravitational wave of very low energy and very low wavelength from the 
mass.  In their Appendix C. Numerical Estimates, Eppley and Hannah attempt to 
show that the device can be built---at least in principle.  As I 
mentioned above, this consists, for their purposes, of showing that 
the universe contains enough mass, is large enough and will last long 
enough.  Here is what is required for the experiment:

\begin{enumerate}
    \item  A (small) mass that will be localized using a 
    gravitational wave.

    \item  A method for giving the mass a small and very well-defined 
    initial momentum.

    \item  A gravitational wave generator to produce short wavelength, 
    low momentum radiation.

    \item  A detector array to measure the new trajectory of the 
    gravitational wave to determine the location of the interaction region.

\end{enumerate}

Requirement 1. is easily enough accommodated.  From the earlier part
of their article, one would assume an uncharged particle, a neutron
for example, should suffice.  Instead, Eppley and Hannah use a $10$ gram mass.  The reason for this is that, as they say, ``[i]f we wish
to keep the detector mass $M_{tot}$ within limits, we need to make the
masses of the generator and the probed particle as large as
possible.''(66)  As I pointed out above, we have no hope of
diffracting a beam of ``particles'' this large.  From their equation
$(A2)$ (see also \cite{mtw}, chs 35-37), we find that the radius of the
detector $R$, is linear in the distance of closest approach and
inversely as the mass of the object to be localized.  If we use a
molecule of mass $\approx 100 m_{p}$ where $m_{p}$ is the proton mass,
we find that, rather than needing a detector radius merely of
approximately $10^{15}cm$ we need instead a radius of order
$10^{37}cm$.  Since the mass of a spherical shell detector goes like
the square of the radius, and their original detector was roughly 1000
galactic masses, we see that the new detector would mass at least
$10^{47}$ galactic masses.  There isn't \emph{this} much mass in the
universe even for the best case estimate.  I will leave this issue
aside for now and return to the detector actually proposed by Eppley
and Hannah.

\subsection{Materials}

The first problem the idealized experimenter faces is one of
materials.  Because the energy of the gravitational wave used to
localize the particle is so low, the wave itself is very difficult to
detect.  Eppley and Hannah require extremely sensitive detectors. 
They use very loosely bound mechanical oscillators---i.e. masses
joined with very weak springs.  How weak?  They never say explicitly. 
We are given that the ground state frequency of the oscillators is of
order $\omega_{0} \sim 10^{-5}sec^{-1}$.  This gives a ratio of
$\sqrt{\frac{k}{m}} \sim 10^{-5}sec^{-1}$, where $k$ is the spring
constant of the oscillator.  For their estimates of the total mass of
the detector array, we can conclude that the mass at each end of the
spring is approximately 1 gram.  But it may be possible to vary this
by an order of magnitude in either direction.  What does this give for
a spring constant?  For $m = 1 gm$, $k = 10^{-10}gm/sec^{2}$.  This
truly is a loosely bound oscillator.  Typical spring constants of the
type described by Eppley and Hannah have $10gm/sec^{2} \leq k \leq
10^{66}gm/sec^{2}$.  The smallest spring constant I know of is for certain extended polymers where $k = 2 \times 10^{-3} gm/sec^{2}$.\cite{meinersquake}  Considering that $c$ and $\hbar$ presumably limit the possible values of spring constants, it is incumbent upon Eppley and Hannah at least to
suggest that their value is possible.  Everything we know about springs suggests not only that their value is not feasible, but that it is not possible {\em even in principle}.\footnote{One might suggest that a free particle is itself a simple harmonic oscillator bound with a spring of $k=0$.  But we need also that average extent of the spring is of order $1cm$.  But a free particle does not have a well-defined equilibrium position (i.e., a position to which it periodically returns after absorbing momentum).  For their argument to be convincing, Eppley and Hannah must show that producing such springs is possible in principle.}

\subsection{Materials}

The springs pose another problem for Eppley and Hannah.  Simply put,
the springs are \emph{too} loosely bound.  Or rather, given how hot it
is in the universe, the springs will be too excited to allow the
idealized experimenter to determine if and when the scattered
gravitational wave has interacted with one of the springs.

The detector array envisioned by Eppley and Hannah is a closely packed
spherical shell of the harmonic oscillators described above.  For the
gravitational wave to be detected, it must interact with one of these
oscillators and excite it in such a manner that its excited state can
be distinguished from its normal operating state.  (The details of the following discussion come from pages 61 and 65.)  We require that the transition time be known with precision of order $\lambda / c \sim 10cm/10^{10}cm/sec = 10^{-9} sec$ in order that the particle's position be measured to accuracy $\Delta x \sim \lambda$.  To measure the final state of the detector we would need a time of order $1/\omega_{0} \sim 10^{5} sec$, much too great for our purposes.    But if the oscillators begin in the ground state we can simply determine whether the state of the detector is above the ground sate, and that allows a much shorter observation time.  Therefore, Eppley and Hannah do not attempt to measure the energy of the
oscillators to great precision.  Rather they measure a broad band of
energies.  The lower edge of this band is above the ground state
energy, but the width of the band is much greater than $\omega_{0}$. 
Thus the time required for the measurement can be made as short as
necessary.  The problem with their argument is in the assumption that
they only need to detect oscillator states above the ground state.  In
fact, virtually all of their detectors will be in highly excited
states.

They calculate the necessary physical characteristics (and the
requisite number) of the oscillators that will allow a probability of
order 1 that the gravitational wave excites one of the oscillators. 
They find (67) that their detectors should have linear dimension of
order 1 centimeter with period of order $10^{5}$ seconds.  This gives
an angular frequency of $\omega_{0} \sim 10^{-5}/sec$.  Standard
quantum mechanical calculations give the ground state energy of such
oscillators as $E_{0} = \hbar\omega_{0} \sim 10^{-20}eV$.  For a
quantum oscillator in a heat bath of temperature $T$, the expectation
value of the energy is given by:
\begin{equation*}
    <E> = \frac{\hbar\omega_{0}}{2} + 
    \frac{\hbar\omega_{0}}{exp(\beta\hbar\omega_{0}) - 1}
\end{equation*}

Here $\beta = \frac{1}{kT}$, where $k$ is Boltzmann's constant.  
Taking $T = 2.7K$, the cosmic background temperature, we find:

 \begin{center}
    $<E> \sim 10^{-20}eV + 
    \frac{10^{-20}eV}{exp(10^{-20}eV/10^{-5}eVK^{-1}2.7K) -1}  \sim 
    10^{-20}eV + \frac{10^{-20}eV}{exp(10^{-15}) - 1} \sim 10^{-20}eV + 
    \frac{10^{-20}eV}{1 + 10^{-15} - 1}  \sim 10^{-5}eV$.
 \end{center} 
Eppley and Hannah give the expectation value of the energy absorbed 
by an oscillator from the gravitational wave as:
\begin{equation*}
    <E_{absorbed}> = 10^{-1} m_{osc}L_{osc}^{2}\hbar
    G/R_{detector}^{2}\lambda^{2}c
\end{equation*}

Their estimate of the density of oscillators at the detector shell is 
$10g/cm^{3}$, so we can take $m_{osc} \approx 10 g$.  The detector 
radius, $R_{detector} \sim 10^{15}cm$, $L_{osc}$ was given above as 
$\sim 1 cm$, and $\lambda \sim 1 cm$.  Thus

\begin{center}
$<E_{absorbed}> \sim \frac{10^{-1} 10g 1
    cm^{2}10^{-16}eVsec10^{-7}cm^{3}/gm\cdot{sec}^{2}}{10^{30}cm^{2}1
    cm^{2}10^{10}cm/sec}  \sim 10^{-63}eV$.
\end{center}

Eppley and Hannah's detector must make the probability 
of exciting \emph{one} of the detectors by \emph{one} increment of 
$\hbar\omega_{0}$ of order unity.  But with the detectors, on 
average, in an energy state $10^{15}$ times the ground state,  we 
can't use their expedient of looking at a wide energy interval the 
lower edge of which is higher than the lowest unexcited state.  For 
here we expect \emph{lots} of the oscillators to be in states higher 
than $<E>$ (around $<E>^{\frac{1}{2}}$ of them).  No 
longer can finding an oscillator above its ground state establish that it has 
registered the gravitational wave.  Instead, we must observe the phase 
characteristics of the oscillators' wave functions.  Now we won't 
be able to detect an interaction on anything like the short time 
scale required for accurate determination of the interaction time, a crucial component of the experiment.  The experiment fails.

Thus at the current temperature of the universe, there is no way to
make the experiment work.  One might suppose that we can simply
refrigerate the (million cubic astronomical unit) region of the
experiment.  This might work.  To make the experiment successful, one
would need to cool the region sufficiently that temperature effects
are unimportant.  The reference Eppley and Hannah use \cite{mtw}
suggests that we would need to make $\hbar\omega_{0} >> kT$ so that
$\frac{\hbar\omega_{0}}{exp(\beta\hbar\omega_{0} - 1)} << 1$.  I.e., $T << 10^{-16}K$.
 Even within the confines of a
thought experiment, such a low temperature over such a large region
seems out of bounds.  Moreover, this refrigerator would have to be fantastically large itself.  Could such a device be built in principle?  Is there time enough and mass enough to build the refrigerator, and allow it to work?  It is doubtful.  Naturally one could wait until the universe
itself cools to a temperature of this order.  There are three problems
with this suggestion however.  First, if the universe is closed---and
thus will ``bounce back'' at some finite time---the universe should
never become this cool.  Second, if the universe ever were at a
temperature of order $10^{-16}K$, there is no indication that enough
free energy would be available to construct Eppley and Hannah's
device.  In any case, we need a concrete argument that shows the
possibility of cooling the experimental region before the experiment
can be considered performable in principle.

There is, moreover, a more conjectural reason to think that temperature will be necessarily too high:  The mass of the device itself may make the springs too hot.  The Unruh effect implies that an observer in a gravitational field will experience a thermal bath of temperature $kT=\hbar a /2\pi c$.  A mass constrained to the surface of a sphere experiences an acceleration given by the surface gravity, since in general relativity the mass is constrained to {\em deviate} from its proper geodesic.  A naive application of the Unruh effect therefore implies that each of the detector oscillators experiences a heat bath due to the mass of the detector array.  For Eppley and Hannah's device, the surface acceleration is $a \sim G \times 1000 m_{galaxy} / 10^{15}cm^{2}$ so at the surface $kT \sim 10^{-33}$.  Thus $T \sim 10^{-17}$, making the device itself too hot for us to perform the experiment, even in principle.

\subsection{Preparation}

To localize the $10 gm$ test particle in momentum space, with its position uncertainty greater than its linear extent, requires an auxiliary experiment.  For this Eppley and Hannah propose to measure very precisely the momentum of a proton, scatter it off the particle, and then measure its momentum again.(pp. 66--67)  Since the test mass is effectively infinite compared to that of the proton, we can take the $x$ component of its velocity to be $v_{fx_{p}} + v_{ix_{p}} = 2v_{x_{tm}}$ where $v_{fx_{p}}$ is the final $x$-velocity of the proton $v_{ix_{p}}$ its intitial $x$-velocity, and $v_{x_{tm}}$ is the $x$-velocity of the test mass.  To determine these quantities within the lifetime of the universe, and within its observable radius (as Eppley and Hannah demand for their experiment) requires a diffraction grating of linear extent $10^{26} cm$ and a measurement time of $10^{17} sec$.  For a larger mass, the uncertainty required in the proton's velocity decreases proportionally to the increase in mass.  The detector extent $W \gtrsim \hbar/(m \Delta v)$ is invariant under changes in test particle mass.  The question of whether there is enough mass in the universe to manufacture such an array, and what effect its mass would have on the remainder of the experimental apparatus is not a negligible issue.  But I will not pursue that question here.  Instead let us focus on their diffraction grating.  The required spacing of the scattering centers is $d\sim 10^{-13} cm$, and this is tightly constrained by cosmological considerations.(p66)  But this is impossible for ordinary matter.  Crystal spacing is of order atomic radius.  For hydrogen $R \approx .37 \mathrm{\AA} = .37 \times 10^{-8} cm$ or approximately $10^{5}$ times too great.  Again it seems incumbent on Eppley and Hannah to show that such a preparation is possible given the materials allowed by the laws of physics.

\subsubsection*{Upshot}

I have offered good grounds to believe that Eppley and Hannah's experiment cannot work.  To convince
us that the experiment could be performed \emph{in principle}, they
would need to provide arguments that show: 1) $c, G$ and $\hbar$
together allow materials with the requisite spring constants; 2) the
entire experiment can be refrigerated sufficiently to allow reliable
detection of the gravitational radiation used in the experiment; 3) appropriate auxiliary measurement protocols could be devised, and the devices to carry them out could be constructed.
Absent these arguments, a \emph{new} device could be
``constructed'', that is, a {\em new} experiment could be performed---one that uses in principle possible materials and takes into
account the temperature of the region of spacetime occupied by the
device.

I cannot comment on the possibility of the latter option---a new
experiment---but I \emph{can} say that the arguments suggested in the
former option---establishing the possibility of the old device---is itself hopeless.  For, even granting, as we should not, the in principle existence of
their ultra-high-tech materials and super-refrigerators,
the experiment cannot (meaningfully) be conducted.  The entire device,
it turns out, sits inside its very own black hole.

\subsection{Gravity Itself}

\subsubsection{Their detector is in a black hole}

The mass of the detector array is sharply constrained by the demands of the experiment:  $R$ must be large enough that the angular resolution set by the detector elements and the radius together is fine enough; the density of the detector elements must be such that there is an appreciable probability of a detection event.  Eppley and Hannah's detector masses approximately $1000 M_{galaxy}$, given by 
$M_{tot} \sim R_{detector}^{3} \times \rho_{oscillators} \sim 10^{15}cm \times 10 gm/cm^{3}$ (p65--66).  Thus its Schwarzschild radius is of order $10^{19}cm$.    Since their estimate of the detector's mean radius is only
$10^{15}cm$, the Schwarzschild radius is $10,000$ times
that of the detector.  But if a mass is contained within its
Schwarzschild radius, it's inside a black hole.  Whatever else we can
say about this experiment, it should be obvious that this is a serious
problem.  How, for example, does one communicate the results with the outside?  Our
experimenters, whatever they observe, are completely cut off from
the rest of the universe.  Moreover, the gravitational stress-energy
is enormous.  The linear approximations
Eppley and Hannah use to derive the sensitivity of their oscillators
cannot hold good in the interior of a black hole.  Further, since null
rays (e.g. gravitational waves) are trapped inside the black hole, the
experimenter would have a real problem establishing the time of
interaction.  Why?  If we suppose that the gravitational wave
propagates out as far as the detector array, we know that it should
return to the array rather than escaping to the exterior of the black
hole.  The question then becomes, ``did it interact with the detector
on the way out or on the way back in?''  Naturally this problem
presupposes that there is still a detector to consider.  Matter that
forms a black hole very rapidly collapses to central region of
extremely high density.  It is then impossible that the
detector, as such, would survive its own construction.

\subsubsection{Possible modifications}

\noindent $1$.  Couldn't we simply enlarge the detector and avoid the whole 
problem of gravitational collapse?  No.  The mass of the detector as a 
whole has to increase at least as fast as the square of the radius.  
Simply put, as the detector radius increases, the surface area 
increases and we need more detectors to ensure a reasonable 
probability of detection.  But the Schwarzschild radius increases 
linearly with the mass, that is, with the square of the detector radius.  So increasing the radius gets us into worse 
shape---the detector is further and further inside its own 
Schwarzschild radius.

\noindent $2$.  What about Eppley and Hannah's suggestion that we can focus the
gravitational radiation?  If we do that, then the total mass is an
upper limit, and need not reflect the ``real'' quantity of mass we
need to use.  The first problem with this suggestion is that to do so
requires even more mass!  To focus gravitational radiation, there's no
substitute for mass.  Putting this issue aside for the moment, we can ask how
much mass we expect to save through focusing.  Notice that using
only a small portion of the detector doesn't help by itself.  For example, if we
focused the radiation into $1\%$ of the detector, then we would need
only $1\%$ of the mass.  But the effective radius of the new detector
(the radius within which all the mass would lie) is also only reduced to
$1\%$ of its earlier value.  But we {\em would} pick up a decrease in the
required mass density if we could focus all of the radiation into the reduced
detector.  For we would need only $100$th as many detectors, the
density of radiation being increased by a factor of $100$.  This is
really the lower limit on how much of the radiation we need to
focus---a factor of $100$ reduction in the mass density would bring us
just outside the Schwarzschild radius.

It is a tricky problem in general relativity to determine if suitable
gravitational lenses could be developed for this application.  The
problem is that, at least for a spherical focusing mass, the angular
deflection of a given ray goes roughly like $M$, the mass of the body,
while the capture cross-section for all rays (i.e. the likelihood of
being absorbed by the lensing matter rather than propagated) goes like
$M^{2}$.  (\cite{waldgr}, 144.)  Thus we reduce
the luminosity of the radiation by absorption faster than we increase
its energy density by focusing.  Without a concrete proposal for how
to focus the gravity waves without damping out the signal, it is not
possible to say that Eppley and Hannah have shown, in principle, how
to keep the mass of the device within physically possible bounds.  Simply claiming that one could focus the radiation does not show that one could really do so.  Even co-ordinating so massive a lens with the rest of the detector---setting it far enough away not to distort the detector, but close enough to produce measurements on usable time scales---is not clearly possible in principle.

Moreover, the $99\%$ reduction in mass density given above isn't
really enough.  As pointed out before, to measure a violation of the
uncertainty principle requires that we observe a beam of particles
that doesn't diffract, even though the individual members of the beam should---and Eppley and Hannah implicitly assent to
this.  Thus we need to reduce the mass-density even further.  Earlier
I suggested a total mass of about $100$ neutrons as the upper limit
for a diffraction experiment.  We will see that even $10,000$ wouldn't
be large enough.  Suppose we reduce the mass of the measured object to
$10,000$ times the mass of the neutron---yielding a mass of about
$10^{-20}gm$.  Then (see Eppley and Hannah's equation C8), the total
mass of the detector (in the absence of focusing) becomes $10^{45}$
galactic masses.  Ignoring the question of whether there is $10^{45}$
galactic masses worth of matter in the universe (there isn't), we need
now to focus not merely $99\%$ of the gravitational radiation into $1\%$ of
the detector, but $\frac{10^{44} - 1}{10^{44}}\%$ of it.  That is, if
we use the same size sector of the detector, and if we wish to avoid gravitational collapse, we must must not introduce more than 10 galactic
masses into that sector.  
A massive enough lens to focus all the radiation into such a small sector is certainly out of the question; a lens that does that without absorbing an appreciable fraction of the beam energy is, apparently, impossible even in principle.

\subsection{Outlook}

Eppley and Hannah claim that their thought experiment would demonstrate 
the inadequacy of any semiclassical theory of gravity.  I have 
outlined two separate objections to that claim, either of which alone is 
sufficient to undermine it.  I have shown that their interpretation 
of the significance of their experiment is untenable and relies on a 
narrow and not unobjectionable interpretation of quantum mechanics.  I have 
further shown that, regardless of how we interpret the results of the 
experiment, the experiment itself cannot be performed---even in 
principle.  Their ``experiment" thus provides no evidence that the 
gravitational field must be quantized.  And thus, one of the most influential attempts to show that gravity is quantized fails.

I have not claimed that no experiment like the one envisioned by 
Eppley and Hannah could be made to work.  However I do not see any way to modify this experiment in order to obtain their results.
And \emph{this} experiment, at least, does not work and cannot be performed---even in principle.  Notice that the problems 
with the experiment are not particularly subtle; their analysis 
requires only basic results in \qm\bs, statistical mechanics, and \gr\bs.  

Perhaps the above discussion will after all prompt the development of a thought experiment that really would show the impossibility of non-quantized gravity.  Such an experiment is likely to suggest directions for constructing a quantum theory of gravity.  Investigating other possible thought-experiments that rule out non-quantized gravity models may also lead to experiments that really {\em can} be performed at some later date.


\begin{thebibliography}{10}
    
\bibitem{ch}
Craig Callender and Nick Huggett, editors (2001), {\em Physics Meets 
Philosophy at the Planck Scale}. Cambridge: Cambridge U. Press.

\bibitem{eh}
K.~Eppley and E.~Hannah.
\newblock ``The Necessity of Quantizing the Gravitational Field.''
\newblock {\em Foundations of Physics}, 7:51--65, 1977.

\bibitem{k}
T.W.B. Kibble.
\newblock ``Is a semiclassical Theory of Gravity Viable?''
\newblock In C.J.~Isham, R.~Penrose and D.W.~Sciama, editors, {\em 
Quantum Gravity 2, A Second Oxford Symposium}, pages 63--80. 
Clarendon: Oxford 1981.
%


\bibitem{jm1}
Mattingly, James (1999), Lecture at the Fifth International Conference on the History and Foundations of General Relativity.  

\bibitem{jm2}
Mattingly, James (2005)
\newblock  ``Is Quantum Gravity Necessary?"
\newblock In Eisenstaedt, Jean, Kox, A.J. editors, {\em The Universe of General Relativity: Einstein Studies}, 325--337.  Boston: Birkh\"auser 2005.


\bibitem{meinersquake}
Jens-Christian Meiners and Stephen R. Quake 
\newblock ``Femtonewton Force Spectroscopy of Single Extended DNA Molecules"
\newblock {\em Physical Review Letters}, 84:5014--5017, 2000.


\bibitem{mtw}
Misner, Charles,  Kip Thorne, and John Wheeler (1973), {\em Gravitation}. New York: W. H. Freeman and Company.

\bibitem{pg}
D.N.~Page and C.D.~Geilker.
\newblock ``Indirect evidence for quantum gravity.''
\newblock {\em Physical Review Letters}, 47:979--982, 1981.

\bibitem{waldgr}
Wald, Robert (1984), \newblock {\em General Relativity}.  Chicago: University of
Chicago Press.
    
\bibitem{waldqft}
Wald, Robert (1994), \newblock {\em Quantum Field Theory in Curved Spacetime and Black Hole Thermodynamics}.  Chicago: University of Chicago Press.
    

\end{thebibliography}
\end{document}